\documentstyle[psfig,aps,prl]{revtex}
\begin{document}
\draft
\title
{Probing the energy bands of a Bose-Einstein
condensate in an optical lattice}

\author
{M.~L. Chiofalo$^\dagger$, S. Succi$^*$
and M.~P. Tosi$^\dagger$}
\address
{$^\dagger$ INFM and Classe di Scienze, Scuola Normale Superiore, 
I-56126 Pisa, Italy\\
$^*$ Istituto Applicazioni Calcolo
``M. Picone'' and INFM, I-00161 Roma, Italy
}
\date{Draft: \today}

\maketitle

\begin{abstract}
We simulate three experimental methods
which could be realized in the laboratory
to probe the band excitation energies and the momentum distribution
of 
a Bose-Einstein condensate inside 
an optical lattice.
The values of the excitation energies
obtained in these different methods agree 
within the accuracy of the simulation. 
The meaning of the results in terms of density and phase 
deformations is tested by studying the relaxation
of a phase-modulated condensate towards the ground state.

\end{abstract}

\pacs{PACS numbers: 67.40.Db, 03.75.-b, 32.80.Pj}

A periodic potential can be imposed on a cold atomic gas from the
shift of the atomic ground state under illumination by a detuned 
laser standing wave \cite{JESSEN}. Experiments on ultracold atoms
inside such an optical lattice have revealed 
dynamical behaviours which are
well known from band theory for electrons in solids, {\it i.e.}
Wannier-Stark
ladders \cite{NIU}, Bloch oscillations \cite{SALOMON}
and Landau-Zener tunneling \cite{RAIZEN}. Especially exciting
perspectives are offered by confinement of an atomic Bose-Einstein
condensate (BEC) inside an optical lattice, in regard to the
generation of coherent matter-wave pulses 
\cite{AK}, laser cooling \cite{CHU}, quantum computing \cite{DIETER}
and more generally BEC quantum-state engineering.

Theoretical and numerical studies of a BEC in a periodic potential
have been based on the Bogolubov-de Gennes equations 
\cite{MOLMER}-\nocite{ZOLLER,ZHANG}\cite{EPJD} and on the 
Gross-Pitaevskii equation (GPE)
\cite{NIU1}-\nocite{PLA,123,PRE}\cite{CARR}.
These studies have shown that, assuming phase coherence in the ground
state, a BEC subject to a 
periodic optical potential and to a constant
external force performs Bloch oscillations as if it were a
quasi-particle inside the lowest energy band. This is consistent with
the observed emission of coherent matter pulses from a vertical array
of optical traps under the force of gravity \cite{AK}.

In this Letter we report the first detailed theoretical analysis of 
the observability of the BEC energy bands under currently 
attainable experimental conditions. 
To this end we use the GPE to simulate three different 
pump-probe experimental methods, which 
have already been realized in the laboratory to measure 
Bloch oscillations of ultracold Bose atoms
\cite{SALOMON} and to study the shape deformation modes 
of a BEC in harmonic traps 
\cite{EXPEXC,EDW}. In interpreting our results it will be useful to
recall that (i) the lowest band arises from pure phase modulations of
the BEC with given $q$-vector in the Brillouin zone, and (ii) 
the higher bands are associated with BEC density profiles having the
symmetry of the isolated-well states, accompanied by
modulations of the phase (see especially Ref. \cite{EPJD}, where
these properties are derived in the Wannier representation for
the wave functions).

We consider a dilute BEC 
in the external potential
$U_l^0[1-\exp(-r^2/r^2_{lb})\cos^2(2\pi z/\lambda)]$, with 
$U_l^0$ being the well depth, $\lambda$ and  
$r_{lb}$ the laser-beam wavelength and waist, 
and $d=\lambda/2$ the lattice period. 
Its dynamics is accurately described by a
one-dimensional (1D)
model, if the mean-field interactions are renormalized to reproduce
the correct 3D value of the chemical potential \cite{123}.  
We thus adopt the 1D GPE for the
BEC wave function $\Psi(z,t)$,
\begin{equation}
i\hbar{\partial\Psi(z,t)\over\partial t}=
\left[-{\hbar^2\nabla_z^2\over 2
M}+U(z,t)
+U_p(z,t)
\right]
\Psi(z,t)\quad .
\label{gpe}
\end{equation}
In Eq. (\ref{gpe}) $U(z,t)$ includes the 1D lattice potential 
$U_l(z)$ and
the mean-field interactions, {\it i.e.} 
$U(z,t)=U_l^0\sin^2(\pi z/d)
+{4\pi\hbar^2 \gamma a\rho}|\Psi(z,t)|^2/M$, with 
$a$ the scattering length, $\gamma$ the renormalization factor, 
$\rho$ 
the number of particles per lattice well and 
$M$ the atom mass. We study three 
different forms for the pump  
potential $U_p(z,t)$, as will be specified below.  
We adopt the system parameters from  
the experiment on $^{87}Rb$ \cite{AK} ($a=110$ Bohr radii,
$\lambda=850\; nm$ and $U_l^0=1.4\;
E_R$ with $E_R=h^2/8Md^2$) and take $\rho=2500$ atoms per well. 

On the technical side, we handle the heavy simulations needed
to obtain significant results on fine spectral structures by 
a well tested explicit-time-marching algorithm \cite{PRE,PREGS}. 
We first determine the ground state by numerically propagating 
Eq. (\ref{gpe}) in imaginary time 
and then insert it as the initial condition for evolution in 
real time. By the same algorithm we preliminarly calculate the band
energies $E_n(q)$ from the relaxation of 
a statically deformed BEC (see below), $n$
being the band index and $q$ the reduced wave vector in the Brillouin
zone.    
The size of the simulation box is 700 wells, as needed 
for a three-digits accuracy 
in the ground-state energy ($\mu=0.695\; E_R$). We 
guard against unwanted localized excitations at
the boundaries by letting the density profile vanish over a length
scale which is much larger than the healing length
$\xi=(8\pi a\rho)^{-1/2}
\simeq 1.4\; d$. Typically, a grid contains $5\cdot 10^4$ points 
and the simulation is carried out up to final times 
ranging from $13$ to $32\; ms$. In the dynamical 
simulations the
data are stored every $20\; \mu s$ and used to
obtain the real and imaginary parts of the Fourier transform of the
wave function ($\Psi(z,\omega)$, say).

We turn to present our numerical results. 
In the first method, which we familiarly refer to as kicking,
we impart a velocity $v=\hbar q/M$ to the BEC at time
$t=0$ by imposing a phase $\hbar qz/M$ on the ground-state 
wave function. The dynamics of the BEC is then monitored
with $U_p=0$.
We show in Fig. 1 the spectra $\Re\Psi(z,\omega)$ 
at any position $z$ in the bulk of the BEC, as functions of
$(\hbar\omega-\mu)/E_R$ for four
values of $q$. At each $q$
the strong peak on the left refers to the lowest band ($n=0$, say) 
and the other two peaks belong to the bands $n=1$ and $n=2$. 
The eigenfrequencies revealed by 
Fig. 1 are the same as those appearing in
$\Im\Psi(z,\omega)$. The density spectra $|\Psi(z,\omega)|^2$  
show instead a strong peak 
centered at energy $\mu$ for all values of $q$ and two
further peaks corresponding to the 
$n=1$ and $n=2$ peaks in Fig. 1, the $n=0$ peak of
Fig. 1 being absent.

\begin{figure}
\centerline{
\psfig{file=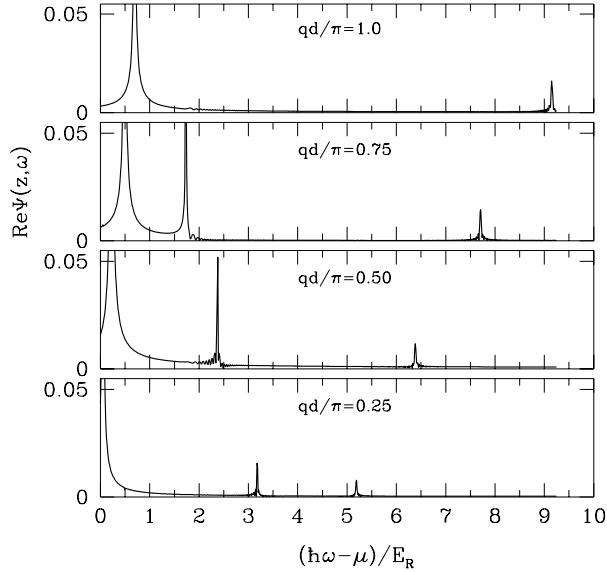,height=8.6cm,width=8.6cm}
}
\caption{ 
Kicking method: bulk spectrum 
$\Re\Psi(z,\omega)$ as a function
of $(\hbar\omega-\mu)/E_R$ for four
values of $q$ in the 
Brillouin zone. 
From bottom to top:
$q=0.25$, $0.50$, $0.75$ and $1.0$ $\pi/d$. 
}
\end{figure}
Thus, a measurement of the density $|\Psi(\Delta z,t)|^2$
of a portion $\Delta z$ of
the BEC by a technique of imaging after ballistic expansion 
\cite{EXPEXC} could reveal excitations in higher bands, but should
have no access to those in the lowest band. This is in accord 
with the aforementioned theoretical result \cite{EPJD} showing that
the excitations in the $n=0$ band correspond to pure phase
modulations. Instead, the momentum distribution 
$n(p,t)=|\int\; dz \Psi(z,t)\exp(ipz)|^2$ contains information 
on the spatial variations of the BEC phase through the 
average velocity. The spectra shown in Fig. 1 could therefore be
experimentally accessible through measurements of the momentum
distribution $n(p,t)$ as a function of the observation time after
imparting an initial velocity (see also the discussion below).

Turning to
the second simulational 
method, that we refer to as shaking, 
we modulate the BEC in space and time
by parametrically driving the lattice potential. To this end
we set $U_p(z,t)=\alpha U_l(z)\cos(qz-\Omega t)$ for $t<t_d$ and
$U_p=0$ otherwise. We choose a small pump amplitude
($\alpha=0.15$), match the pump frequency $\Omega$ near resonance
and tune the drive-time $t_d$ over several excitation periods, the
time unit being $T_R\equiv h/E_R\simeq 0.32\; ms$.
The density is then recorded at times $t\gg t_d$.

Fig. 2 displays the bulk density spectrum 
$|\Psi(z,\omega)|^2$  
for the same $q$-values as in Fig. 1. The spectrum contains a
peak centered at energy $\mu$ for all values of $q$ and 
a strong peak corresponding to the
nearly resonant drive, at a frequency changing with $q$ 
in accord with the $n=0$ peaks in Fig. 1.
The other structures in Fig. 2 can be identified as
harmonics of the fundamental excitation energy and combinations of
them.

Thus a near-resonance measurement of the density $|\Psi(z,t)|^2$
in the shaking method may 
reveal 
the excitations in all bands and in particular those in the $n=0$ band.
Shaking at $q=0$ 
coherently drives the bound states in each 
well and creates pure density fluctuations. 
Phase modulations are also triggered at $q\neq 0$, due to the  
spatial periodicity of the BEC: this is
at variance from
the shape-deformation modes in harmonic traps \cite{EDW}. 
The energies of these phase and
density excitations are located at the poles 
of the response function 
to the
external drive and therefore 
show up as peaks in $|\Psi(z,\omega)|^2$.  

\begin{figure}
\centerline{
\psfig{file=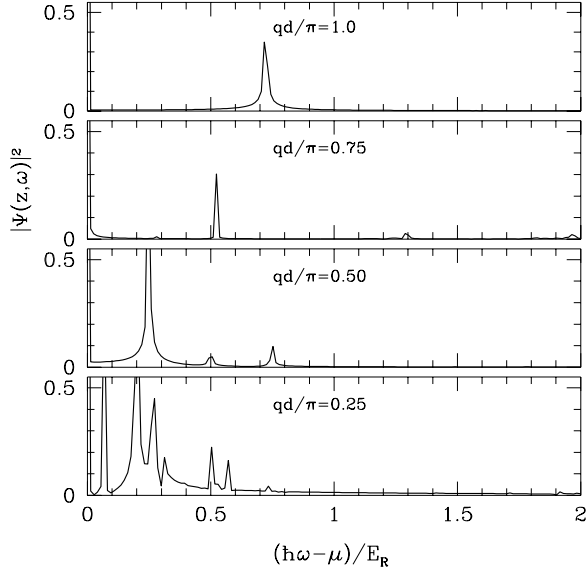,height=8.6cm,width=8.6cm}
}
\caption{
Shaking method: bulk density
$|\Psi(z,\omega)|^2$ as a function
of $(\hbar\omega-\mu)/E_R$ for the same $q$-values 
as in Fig. 1.
}
\end{figure}

Before quantitatively comparing the results obtained by the above
dynamical methods, we pause to present the band structure that we have
obtained by a static deformation 
method. In this method we imprint a phase modulation on the BEC
ground state so as to generate a state having some
overlap with a Bloch state of
quasi-momentum $q$. 
We then propagate this modulated BEC in imaginary time and
monitor the average energy $<E>$ as the BEC returns to its ground state. 
Fig. 3 shows $<E>$ vs. $it\hbar/E_R$ for the
same $q$-values as in Fig. 1.  
Two plateaus are met during this evolution: the first lies 
at energy $E_0(q)$ and the second at energy $\mu$. 
The time scales for the appearance of the plateaus 
correspond to interband and intraband 
relaxation, respectively.
Similar results are obtained 
for the higher bands $E_n(q)$ from phase modulations having wave
vector $q+n'\pi/d$ outside the first Brillouin zone (with $n'=n$ for
even $n$ and $n'=n+1$ for odd $n$).
\begin{figure}
\centerline{
\psfig{file=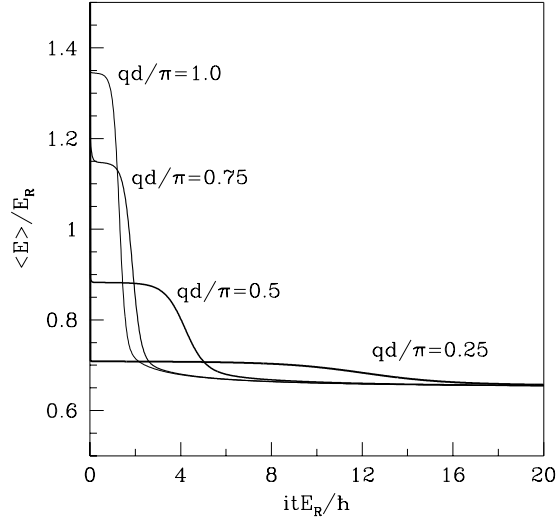,height=8.6cm,width=8.6cm}
}
\caption{Static deformation method:
average energy $<E>$ vs. imaginary time $it\hbar/E_R$ 
for a modulated BEC with the same $q$
values as in Figs. 1 and 2.
Higher plateaus correpond to higher values of $q$ in the lowest 
band.
}
\end{figure}

Fig. 4 reports the energy bands $E_n(q)$ that we have obtained by the
three methods presented above, for $n=0$, $1$ and $2$. The three
methods yield the same excitation energies within 
the simulation accuracy, which is reflected by the size of the symbols
(squares for kicking, circles for shaking and
triangles for the static method) in
the main body of the figure.  
A gap has opened at $q=\pi/d$ between the $n=0$ and $n=1$ bands,
while there is no gap at $q=0$ between the $n=1$ and $n=2$ bands: this 
is a well known result for $1D$ lattice potentials having a doubled period 
\cite{SLATER}. We also remark that the results shown in Fig. 4
for the $n=1$ and $n=2$ bands 
do not differ significantly  from those that we obtain for a non-interacting BEC:
indeed, for the present system parameters we have $U_l^0\simeq 2.12\mu$, 
implying that 
mean-field interactions are weak on the energy scale of the lattice.

\begin{figure}
\centerline{
\psfig{file=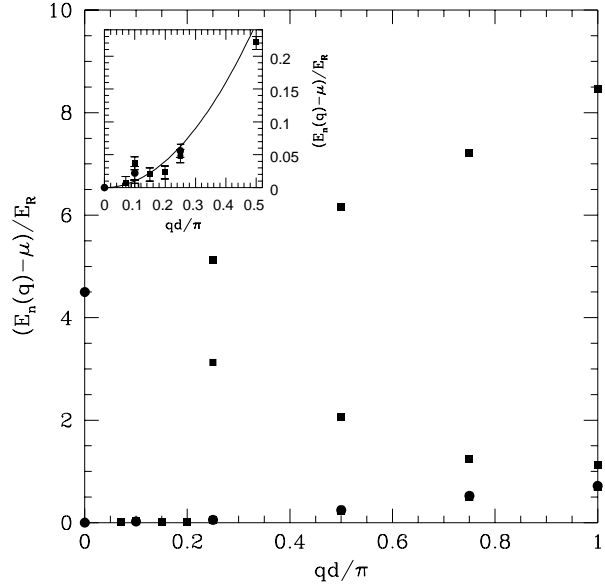,height=8.6cm,width=8.6cm}
}
\caption{The first three 
energy bands in the Brillouin zone $0\leq q\leq
\pi/d$ as obtained by the
methods reported in Figs. 1-3 (squares: kicking; 
circles: shaking; triangles: static method). 
The inset zooms on the lowest band. The size of the error bars
reflects a two-digit accuracy in energy differences. 
Solid line: quadratic dispersion with the bare atom mass.
}
\end{figure}

The inset in Fig. 4 shows an enlarged view of the
excitation energies in the lowest band up to $qd/\pi=0.5$, where the
band starts bending over. On this scale the estimated error bars of
the simulation become visible and well within them the results 
for the present interacting BEC agree with those for the
non-interacting one, except perhaps at $qd/\pi=0.1$. The calculated
band energies are compatible with the free-particle dispersion
relation $\hbar^2q^2/2M$ as shown by the solid line in the inset. High
resolution and fine tuning of the strength of the self-interactions
relative to the height of the lattice confinement would clearly be
needed to experimentally reveal the phonon-like 
dispersion relation at
long wavelengths which is predicted by gapless theories 
\cite{MOLMER,EPJD}. From the Bogolubov dispersion relation this linear behaviour
may be expected to become visible below 
$q\approx\sqrt{2}/\xi$. 

Finally, we return to discuss the momentum
distribution $n(p,t)$ for a BEC in an optical lattice. We simulate a
feasible experimental method for its measurement by driving the BEC
with a force $F$ for a variable amount of time $t_d$. The force is
taken as a positive constant in the range 
$0\leq t_d\leq T_B/2$ and as a negative constant in the range $T_B/2<t_d\leq T_B$,  
with $T_B=h/|F|d$ being the period of Bloch
oscillations.
We set $U_p(z,t)=-Fz$ for $t<t_d$ and $U_p=0$ otherwise, 
and choose the value of $F$ to correspond to a particle acceleration
of $85 \; cm/s^2$ as in the experiments of Ben Dahan {\it et al.} \cite{SALOMON}. 
The Brillouin zone is explored in this method according to 
$qd/\pi=2sgn(F)t_d/T_B$. 

Fig. 5 shows $n(p,t)$  at times $0\leq
t_d\leq T_B$ covering one 
full Bloch oscillation. The peak seen at $t_d=0$ drifts
to the right up to the first zone boundary at $t_d=T_B/2$,
while a second peak emerges in the second zone and moves to the left
entering the first zone at later times. 
The sharpness of the peaks shows that 
the BEC is behaving as if it were a quasi-particle reflected
back and forth at the Bragg planes. The inset in Fig. 5 shows that the
BEC average velocity $<v>$, as obtained from the momentum distribution
after backfolding of the second zone, is consistent with the
semiclassical average velocity as calculated from $dE_0(q)/dq$ using
the results in Fig. 4.
\begin{figure}
\centerline{
\psfig{file=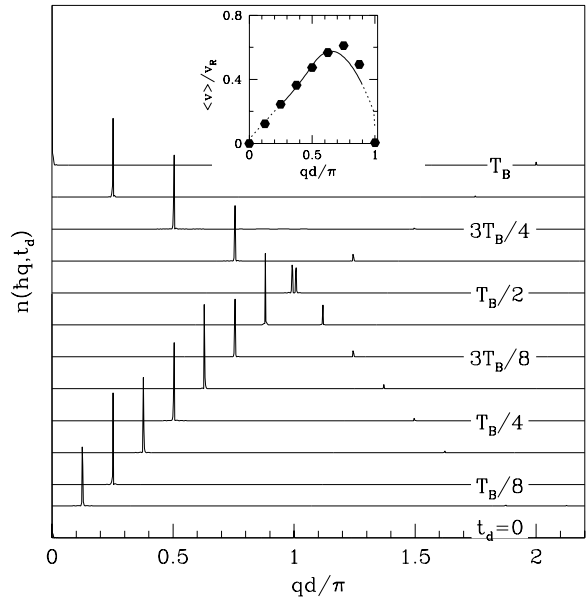,height=8.6cm,width=8.6cm}
}
\caption{Force-driving method:
momentum distribution (in arbitrary units) at various
times $t_d$, equally spaced by $T_B/16$ up to $T_B/2$ and then
by $T_B/8$ up to $T_B$. 
Inset: points give the BEC average velocity
$<v>$ (in units of $v_R=\hbar\pi/Md$) 
as determined from the momentum
distribution after backfolding of the second Brillouin zone, while 
the line shows the semiclassical average velocity.}
\end{figure}

In conclusion, we have simulated three different experimental methods to probe the 
energy bands of a condensate in an optical lattice and tested their
meaning in terms of the density and phase deformations associated with
excited band states.
The kicking and the
shaking methods do not require any special modelling of the
observables in an actual experiment. The force-driving method is
especially suited for a measurement of the BEC momentum distribution
in an optical lattice. A study of 
the observability of the expected linear dispersion 
at small $q$-values would require different system parameters 
than the present ones and in particular 
stronger mean-field interactions and larger lattice 
constants. Theoretical and numerical work is in progress 
to explore these possibilities. 

One of us (MLC) thanks
Dr J.H. M\"uller
for enlightening discussions on experimental issues. Part of this work was 
done using the computational facilities of Cineca. We acknowledge
support from MURST through PRIN2000.

\end{document}